\documentclass[reprint, amsmath,amssymb,aps,prd]{revtex4-2}
\usepackage{subfigure}
\usepackage{graphicx}% Include figure files
\usepackage{dcolumn}% Align table columns on decimal point
\usepackage {cancel}
\usepackage{ulem}
\usepackage[colorlinks,urlcolor=aa,linkcolor=blue,anchorcolor=aa,citecolor=aa]{hyperref}
\usepackage{orcidlink}
\usepackage{bm}% bold math
\usepackage{xcolor}
\usepackage[mathlines]{lineno}% Enable numbering of text and display math
\usepackage{multirow}
\usepackage{enumerate}
\usepackage{amsmath}

\usepackage{lineno}
\usepackage{mathtools}
\usepackage{overpic}
\usepackage{epsfig}
\usepackage{rotating}
\usepackage{longtable}
\usepackage{color}
\usepackage{siunitx}

\definecolor{aa}{RGB}{0,0,139}

\newcommand{\gev}{\rm GeV}

\newcommand{\gevcs}{{\rm GeV}/c^2}
\newcommand{\mev}{\rm MeV}
\newcommand{\mevcs}{{\rm MeV}/c^2}

\newcommand{\g}{\gamma}

\newcommand{\rr}{\rho\rho}
\newcommand{\XiXi}{\Xi^{-}\bar{\Xi}^{+}}
\newcommand{\KKpiz}{K^{+}K^{-}\pi^{0}}
\newcommand{\etac}{\eta_{c}}

\newcommand{\psp}{\psi(3686)}
\newcommand{\jpsi}{J/\psi}

\newcommand{\Eg}{E_{\gamma}}
\newcommand{\Egt}{E_{\gamma}^{3}}

\newcommand{\reduline}{\bgroup\markoverwith
{\textcolor{red}{\rule[0.5ex]{2pt}{0.4pt}}}\ULon}

\newcommand{\beq}{\begin{equation}}
\newcommand{\eeq}{\end{equation}}
\newcommand{\beqar}{\begin{eqnarray}}
\newcommand{\eeqar}{\end{eqnarray}}
\newcommand{\bitm}{\begin{itemize}}
\newcommand{\eitm}{\end{itemize}}

\def\NP{Nucl. Phys.}

\def\PLB{Phys. Lett. B}
\def\PRL{Phys. Rev. Lett.}
\def\PRD{Phys. Rev. D}

\def\ZPC{Z. Phys. C}

\def\RPP{ Rep. Prog. Phys. }
\def\RMP{Rev. Mod. Phys. }

\def\JHEP{J. High Energ. Phys.}

\begin{document}

%\preprint{APS/123-QED}

\title{{\bf \boldmath Line shape of the $\jpsi \to \g \eta_{c}$ decay}}
\author{Ting Wang$^{1}$\hspace{-1.5mm}$^{~\orcidlink{0009-0009-5598-6157}}$, Xiaolong Wang$^{1}$, Guangrui Liao$^{2}$, Kai Zhu$^{3}$
%\author{Ting Wang$^{1}$, Xiaolong Wang$^{1}$, Guangrui Liao$^{2}$, Kai Zhu$^{3}$
\\
\vspace{0.2cm} {\it
$^{1}$ Institute of Modern Physics, Fudan University, Shanghai 200433, People's Republic of China\\
$^{2}$ School of Physics Science and Technology, Guangxi Normal University, Guilin 541004, People's Republic of China\\
$^{3}$ Institute of High Energy Physics, Chinese Academy of Sciences, Beijing 100049, People's Republic of China\\
}}
\vspace{0.4cm}
\date{\today}

\begin{abstract}
An accurate description of the photon spectrum line shape is essential for extracting resonance parameters of the
$\eta_c$ meson through the radiative transition $J/\psi \to \gamma \eta_c$. However, a persistent challenge remains in
the form of a divergent tail at high photon energies, arising from the $E_{\gamma}^3$ factor in theoretical
calculations. Various damping functions have been proposed to mitigate this effect in practical experiments, but
their empirical nature lacks a rigorous theoretical basis. In this study, we introduce two key considerations:
incorporating full-order contributions of the Bessel function in the overlap integral of charmonium wave functions and the
phase space factor neglected in previous experimental studies. By accounting for these factors, we demonstrate
a more rational and effective damping function of the divergent tail associated with the $E_{\gamma}^3$ term. We
present the implications of these findings on experimental measurements and provide further insights through
toy Monte Carlo simulations.
\end{abstract}

%\pacs{Valid PACS appear here}
\maketitle

%\tableofcontents
\section{Introduction}
\label{sec:introduction}

Although Quantum Chromodynamics (QCD), a gauge field theory describing the strong interaction, has been successfully
validated in the high energy regime. However, it continues to face unresolved challenges in the non-perturbative
domain, i.e., at low energies. Charmonium states, whose masses straddle the boundary between perturbative and
non-perturbative regions of the strong interaction, have emerged as a vital testing ground for exploring these
complexities since the discovery of the $\jpsi$ meson fifty years ago. Various measurements, such as the masses
and widths of these resonances, the transition rates, and the decays to light hadron states, have gained valuable
insights into interactions within this energy spectrum. In particular,
precise measurement of the mass and width of the lowest lying $S$-wave spin singlet charmonium state $\etac$ holds
significant importance for advancing our understanding of the strong interaction. For the measurement, a precise
and accurate description of the line shape in the radiative transition $\jpsi \to \gamma \etac$ is essential.

The $\etac$ meson is typically produced in $B$-meson decay, $\gamma \gamma$ collision, or radiative transitions
from $\jpsi$ or $\psp$ state, etc. In the determination of its resonant parameters, a study by Ref.~\cite{Segovia_23}
found that analyzing the same data set with different fitting functions can yield varying mass and width values. Therefore, the uncertainty in the line shape contributes to significant systematic uncertainty in the relevant
measurements. Theoretical investigations into the line shape have been conducted using potential
models~\cite{Eichten1_5, Eichten2_7, Zambetakis_8, Sucher_13, Kang_9, Fayyazuddin_10, lahde_15, lahde_16, Grotch_12,
Feinberg_14, Grotch_17, Ebert_18, Barnes_19, vairo_21, Pineda_22, Segovia_23, Brambilla_24, Brambilla_25, Zhangxg_28,
Deng_31}, lattice QCD~\cite{Dudek_32, Gui_34, Cheny_35, Becirevic_29, Donald_30, HPQCD_20}, and sum
rules~\cite{Khodjamirian_6, shifman_2, Beilin_3, Guosp_27}. In the transition of $\jpsi\to\gamma\etac$, a common
factor $\Eg^3$ is present in the formulas of all these theoretical frames for its partial width, where $\Eg$ is the
energy of the radiative photon. However, this factor exhibits divergence as $\Eg$ approaches a high energy region,
rendering the description unreliable and conflicting with experimental observations. To our best knowledge, no
theoretical mechanism has yet been proposed to address this divergence issue. Instead, empirical damping functions
have been introduced as a solution in previous experimental measurements, such as CLEO~\cite{CLEO} and
KEDR~\cite{KEDR}. Lacking a solid theoretical foundation for these empirical damping functions raises
questions in the accuracy and precision of the experimental measurements. Therefore, developing a theoretically
grounded damping function is desired to accurately describe the line shape of the radiative transition and precisely
extract the resonant parameters of $\etac$.

In this article, we revisit the transition $\jpsi \to \gamma \etac$ using the non-relativistic potential model. We
find that including the full-order contributions of the Bessel function in the overlap integral of charmonium wave functions
effectively mitigates the divergence associated with the $\Eg^3$ factor.  
Additionally, by incorporating the previously overlooked phase space factor in experimental analyses, we introduce a novel theoretically grounded damping function.
To elucidate their impacts on experimental measurements, we conduct toy Monte Carlo (MC)
simulations and offer numerical results for specific $\etac$ decay channels.

\section{framework, calculation, and results}
\label{sec:method}

The following calculation in this article will generally follow the schemes of the potential models. At leading order,
the non-relativistic QCD predicts that the magnetic dipole (M1) amplitudes between two heavy $S$-wave
quarkonia are independent of the potential model. The spatial overlap matrix element is always $=1$ for states
within the same multiplet that contains states with the same radial quantum number, and $=0$ for allowed transitions between different multiplets. With the relativistic corrections due
to spin relevance included in Hamiltonian, the M1 amplitude between an initial state $i=n^{2s+1}L_{J}$ and a final
state $f=n^{\prime ~2s^{\prime}+1}L_{J^{\prime}}$ ($L=0$) can be calculated by~\cite{Eichten:2007qx}
\begin{equation}\label{eq_2}
\Gamma(i \xrightarrow{M_{1}} \g+f) = \\
\frac{4}{3} \alpha e_q^2 \frac{\Eg^3}{m_q^2}\left(2 J^{\prime}+1\right) \left|\mathcal{M}_{i f}\right|^2\ ,
\end{equation}
where $\alpha$ is the fine structure constant, $e_q$ and $m_q$ are the electrical charge and the mass of the heavy
quark $q$, $m_i$ ($m_f$) is the mass of the initial (final) state quarkonia and $\Eg = (m_i^2-m_f^2) /(2 m_i)$.

The matrix element $\mathcal{M}_{if}$ is given by the overlap integral of the wave functions of the charmonia, i.e.,
\begin{equation}
\label{over_lap}
%\begin{aligned}
 \mathcal{M}_{if} = (1+\kappa_q)
\int_0^{\infty} u_{nl}(r) u_{n^{\prime}l}^{\prime}(r) j_0(\frac{\Eg r}{2}) dr,
%\end{aligned}
\end{equation}
where $\kappa_q$ is the anomalous magnetic moment of a heavy quarkonium $q\bar{q}$, $r$ is the relative distance
between $q$ and $\bar{q}$, $u_{nl}(r)$ and $u_{n^{\prime}l}^{\prime}(r)$ are the radial wave functions of the initial
and final states,  and $j_0(x)$ is the spherical Bessel function of the first kind. For the study of $\jpsi\to \g\etac$ transition, $\kappa_q$ is set to zero, and the $u_{nl}(r)$
function of $\jpsi$ and $u_{n^{\prime}l}^{\prime}(r)$ function of $\etac$ are obtained by a specific potential model.
We adopt the non-relativistic potential model~\cite{Barnes:2005pb}
\begin{equation}
V_0(r) = -\frac{4\alpha_s}{3r} + br,
\end{equation}
plus spin-dependent term
\begin{equation}
V_s = \frac{32\pi\alpha_s}{9m^2_c} \delta_\sigma (r) \vec{S}_c \cdot \vec{S}_{\bar{c}} + \frac{1}{m^2_c}
[(\frac{2\alpha_s}{r^3} - \frac{b}{2r}) \vec{L} \cdot \vec{S} + \frac{4\alpha_s}{r^3} T],
\end{equation}
for the spin-spin, spin-orbit, and tensor interactions. Here, $\delta_\sigma(r)=(\sigma/\sqrt{\pi})^3e^{-\sigma^2r^2}$
is a Gaussian-like function that smears the contact term. The four parameters $\alpha_s = 0.54$, $b = 0.15$, $m_c =
1.47~\gevcs$, and $\sigma = 1.05$ are determined by fitting to the world-averaged values of the charmonium
masses~\cite{PDG}. Furthermore, we take the expansion of spherical Bessel function
\beq
j_{0}(\frac{\Eg r}{2}) \equiv \sum_{k = 0}^{+\infty}
\frac{(-1)^k}{k!\Gamma(k+1)}\left(\frac{\Eg r}{4}\right)^{2k}.
\eeq

\begin{figure}[!htbp]
\centering
\begin{overpic}[angle=0,width=0.45\textwidth]{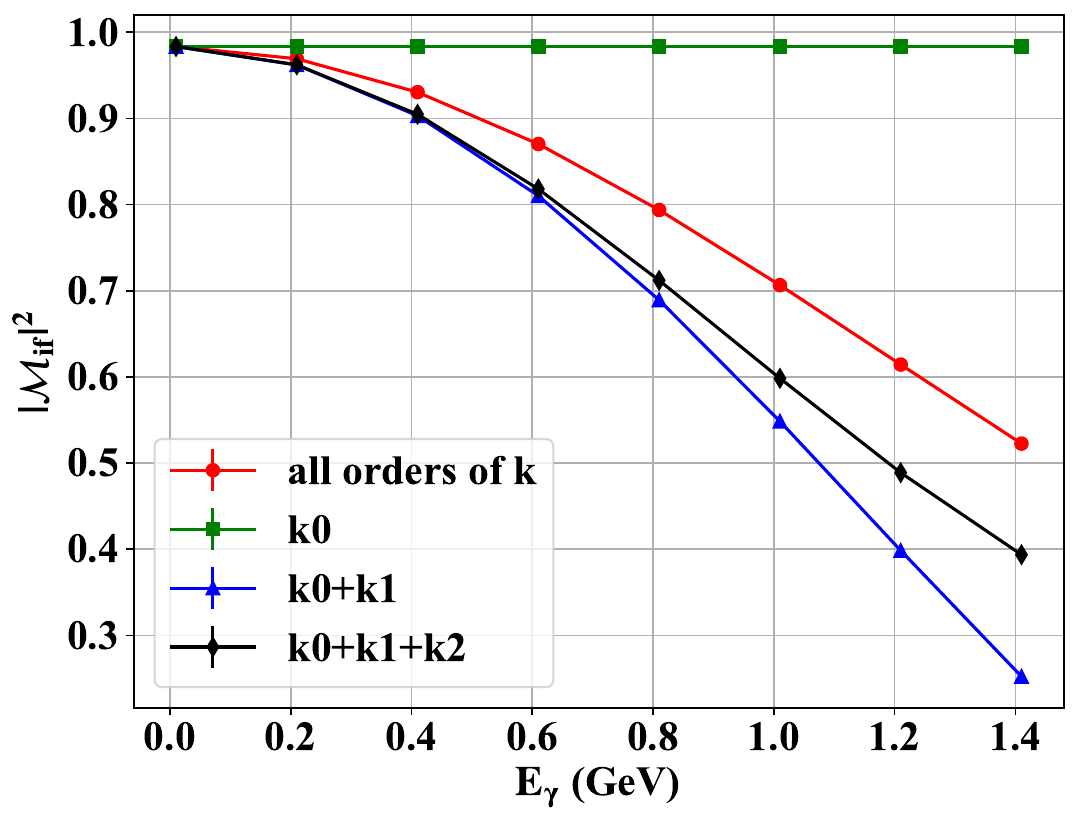}
\end{overpic}
\caption{The $|\mathcal{M}_{if}|^2$ overlap integrals. The green square curve, the blue trigonometric curve, the black
diamond curve, and the red circle curve show the leading order of the Bessel function expansion, the order till the
second, the order till the third, and the full order.}
\label{Fig:overlap_int}
\end{figure}

In previous works such as Refs.~\cite{shifman_2,Eichten2_7,Brambilla_24},  only the leading order of the Bessel
function $j_{0}(\frac{\Eg r}{2})$ is adopted for simplicity. This approximation is fine when the $\Eg$ is
low. However, a high value of $\Eg$ that can achieve even more than $1~\gev$ in some decay channels will result in
a divergence in the partial width formula. Figure~\ref{Fig:overlap_int} shows the $|\mathcal{M}_{if}|^2$ with Bessel
function containing different orders, and it is clear that a suppression emerges significantly with the increase of
$\Eg$ when high order terms of the Bessel function are included. To apply this result to the experimental
measurements, we fit the $|\mathcal{M}_{if}|^2$ with full order contributions of the Bessel function by a polynomial
function and obtain the damping function to be
\begin{equation}\label{EqD}
D(\Eg) = 0.11\Eg^3 - 0.40\Eg^2 + 0.019\Eg + 0.98.
\end{equation}
This is different to the one mentioned in Ref.~\cite{Dudek_32}, which has $|\mathcal{M}_{if}|^2 \propto \exp(-\Eg^2/16
\beta^2)$ with $\beta = 540 \pm 10~\mev$.

\begin{figure}[!htbp]
\centering
\begin{overpic}[angle=0,width=0.45\textwidth]{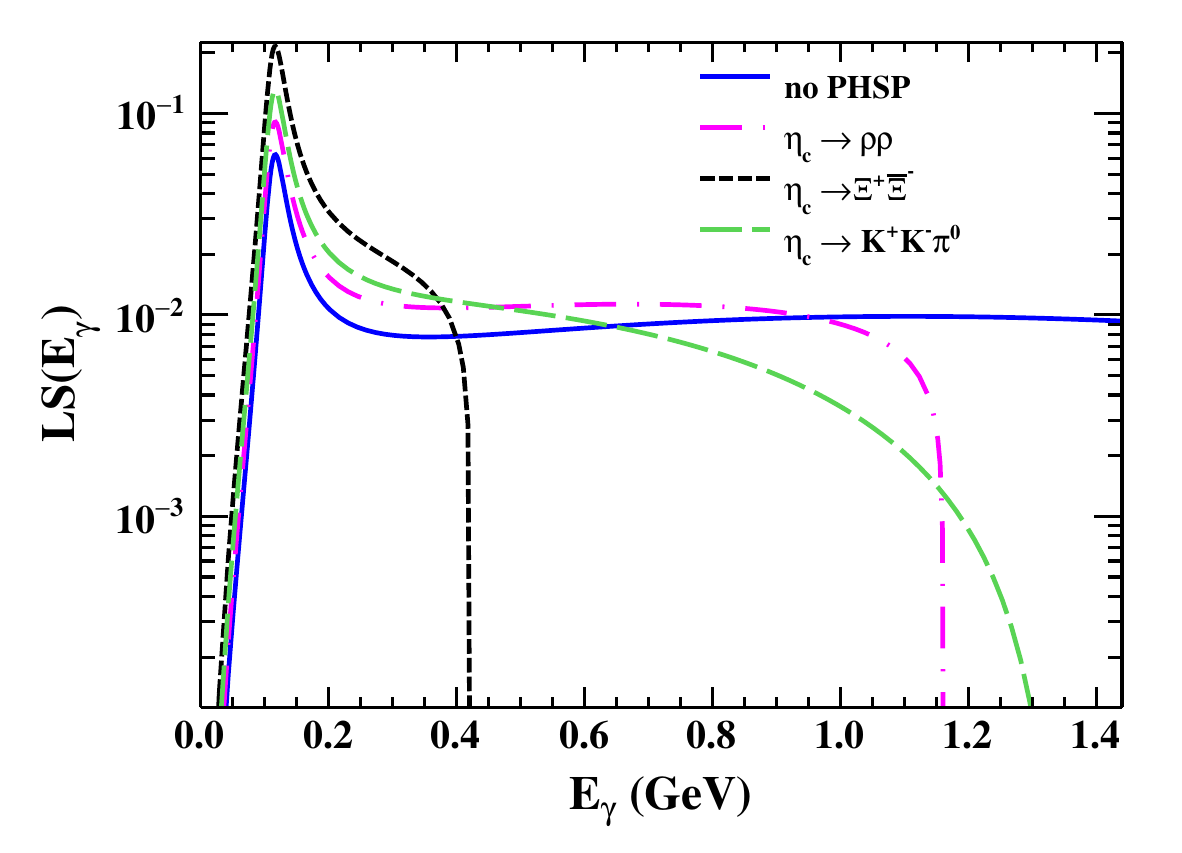}
\end{overpic}
\caption{The line shapes of $\jpsi \to \g \etac$ decays. The blue solid curve contains no phase space factor; the
dashed curves with the color pink, black, and green contain the phase spaces of $\etac \to \rr$ decay, $\etac \to
\XiXi$ decay, and $\etac \to \KKpiz$ decay, respectively.}
\label{Fig:comp_M2_PHSP}
\end{figure}

The phase space of the $\etac$ decay is another factor usually ignored in experimental measurements and
theoretical calculations. The phase space can be approximated as a constant when the invariant mass of the final
states is around the peak of the $\etac$ mass. But it also can decrease significantly when the $\Eg$ increases to a large value. In some channels, this decrease is very fast. Therefore,
our results for the line shape of $\jpsi \to \gamma \etac$ decay is described as
\begin{equation}
\label{eq:normal_line_shape}
LS(\Eg) = \Egt \times BW(\Eg) \times D(\Eg) \times \frac{P(\Eg)}{P(\Eg^{\rm peak})},
\end{equation}
where $BW(\Eg)$ is the Breit-Wigner function
$$BW(\Eg)=\frac{\Gamma_{\etac} / 2\pi}{\left(M_{\jpsi}-M_{\etac}-\Eg \right)^{2}+\Gamma_{\etac}^{2} / 4}$$
describing the $\etac$ resonance, $D(\Eg)$ is the damping function obtained in Eq.~\ref{EqD}, $P(\Eg)$ is the phase space function depending on the final states~\cite{PDG}, $\Eg^{\rm peak}$
is the $\Eg$ corresponding to the peak of the $\etac$ mass. Figure~\ref{Fig:comp_M2_PHSP} shows the line shapes with
phase spaces of $\etac\to \rr, ~\XiXi$, and $\KKpiz$ decays, comparing to the one takes the  constant of
$P(\Eg^{\rm peak})$. There are substantial suppression effects due to the phase space, which varies for different
decay channels.

In short, to determine the line shape of the $\jpsi\to\g\etac$ transition accurately
and precisely, a new damping function considering the high order contributions in the overlap integral of the $\jpsi$ and $\etac$ wave functions is proposed. By incorporating the new damping function and the phase space factor, the final line shape of the $\jpsi\to\g\etac$ decay is given by Eq.~(\ref{eq:normal_line_shape}).
 
%%%%%%%%%%%%%%%%%%%%%%%% SECTION II  %%%%%%%%%%%%%%%%%%%%%%%%%%%%%%%%%%%%%
\section{numerical simulation}
\label{sec:simulation}

\begin{figure}[htbp]
\centering
\begin{overpic}[angle=0,width=0.45\textwidth]{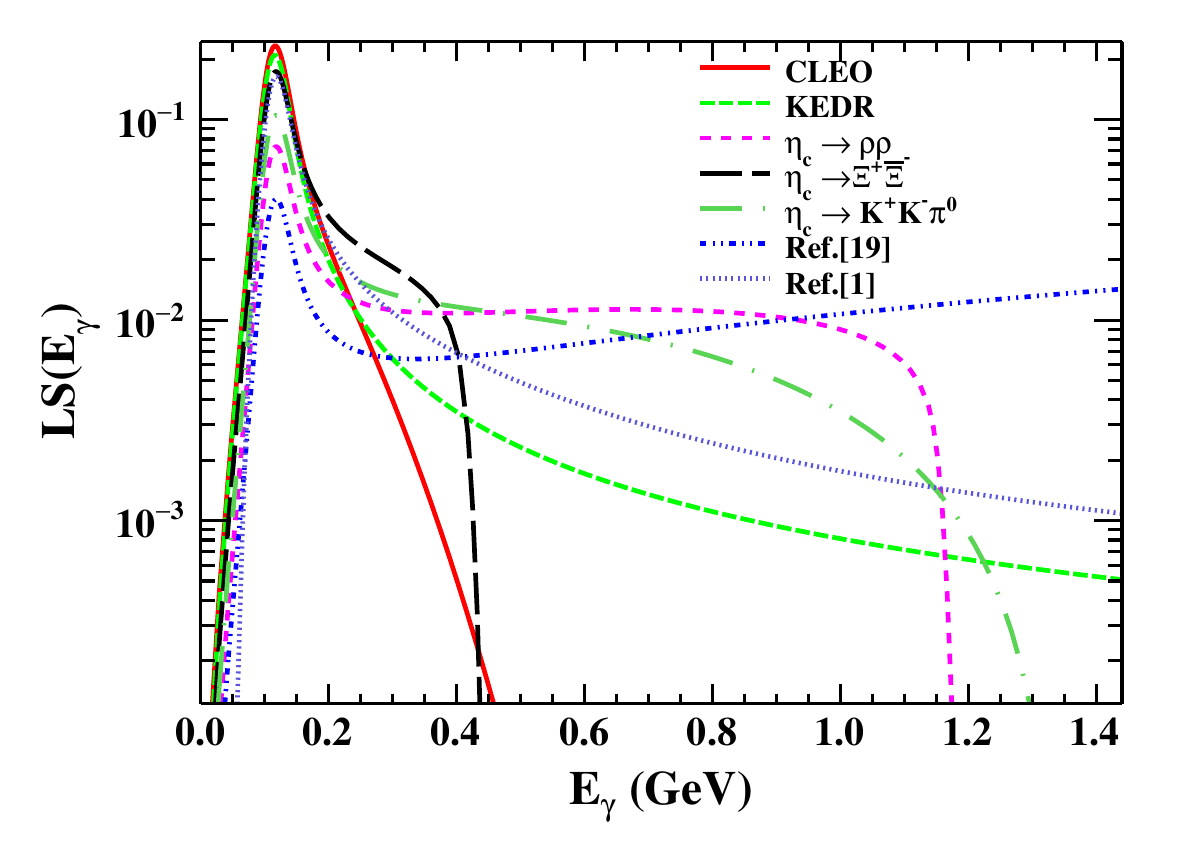}
\end{overpic}
\caption{The $\jpsi \to \g \etac$ line shapes with varied damping functions. The red solid line indicates CLEO's
\cite{CLEO} damping function, and the green dashed line indicates KEDR's \cite{KEDR} damping function. Our damping
functions for $\etac \to \rr$, $\etac \to \XiXi$, and $\etac \to \KKpiz$ decays are presented in pink dashed line,
black long dashed line and dark green short dashed line, respectively. The blue and purple dashed lines are based
on the theoretical calculations in Ref.~\cite{Brambilla_25} and Ref.~\cite{Segovia_23}, respectively. }
\label{Fig:com_line_shape}
\end{figure}

To further study the implications of the two considerations, we compare the newly obtained damping functions with those
used in previously experimental measurements. Two of the most widely used damping functions are $D_{\rm CLEO}(\Eg) =
\exp(-\Eg^2/8\beta^2)$ with $\beta = -65.0 \pm 2.5~\mev$ from the CLEO experiment~\cite{CLEO} and $D_{\rm KEDR}(\Eg) =
(\Eg^{\rm peak})^2/[\Eg \Eg^{\rm peak} + (\Eg - \Eg^{\rm peak})^2]$ from the KEDR experiment~\cite{KEDR}. We also
study two line shapes based on theoretical calculations~\cite{Segovia_23, Brambilla_25}, as discussed before, they all diverge with the increase of $\Eg$. To make the line shapes more
realistic, all of them are smeared by a Gaussian function with a resolution of $10~\mev$. They are displayed in
Fig.~\ref{Fig:com_line_shape} for comparison.

\begin{figure*}[!htbp]
\centering
\mbox{
\begin{overpic}[angle=0,width=0.32\textwidth]{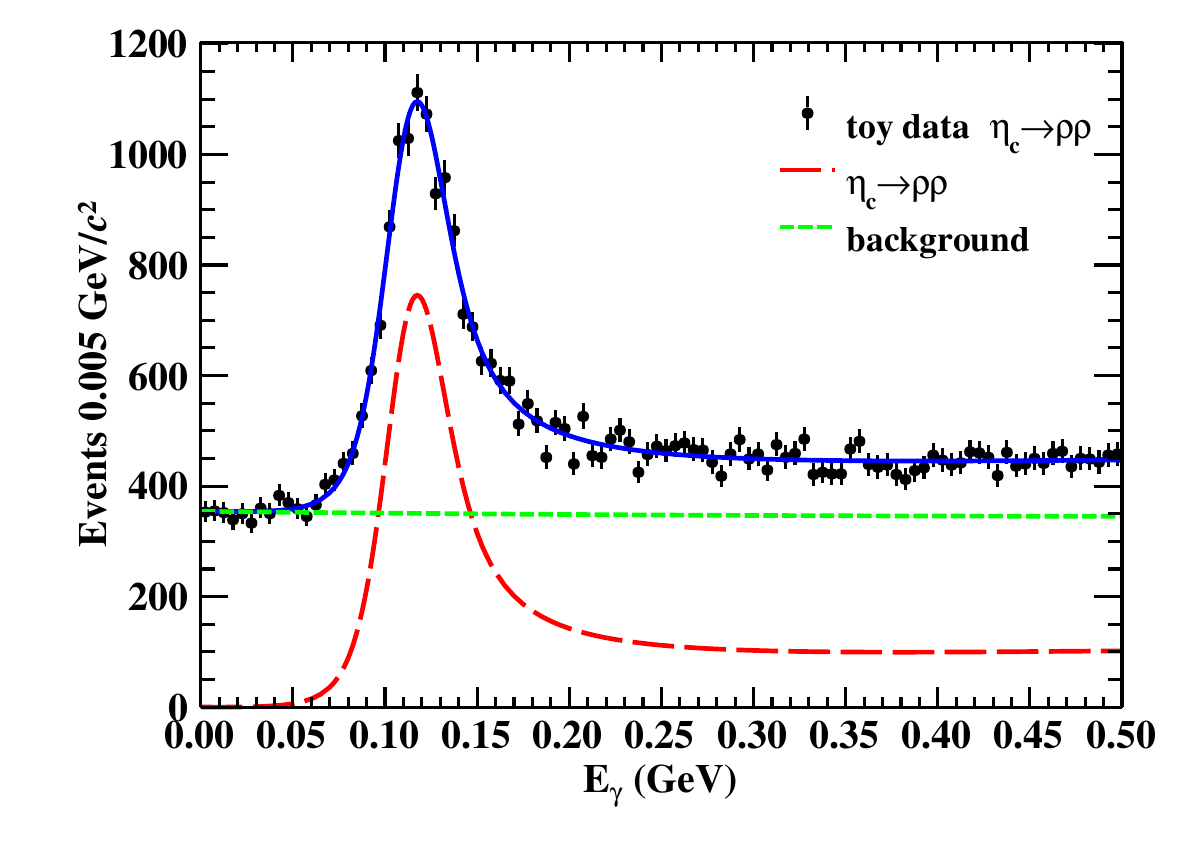}
\put(20,60){$(a)$}
\end{overpic}
\begin{overpic}[angle=0,width=0.32\textwidth]{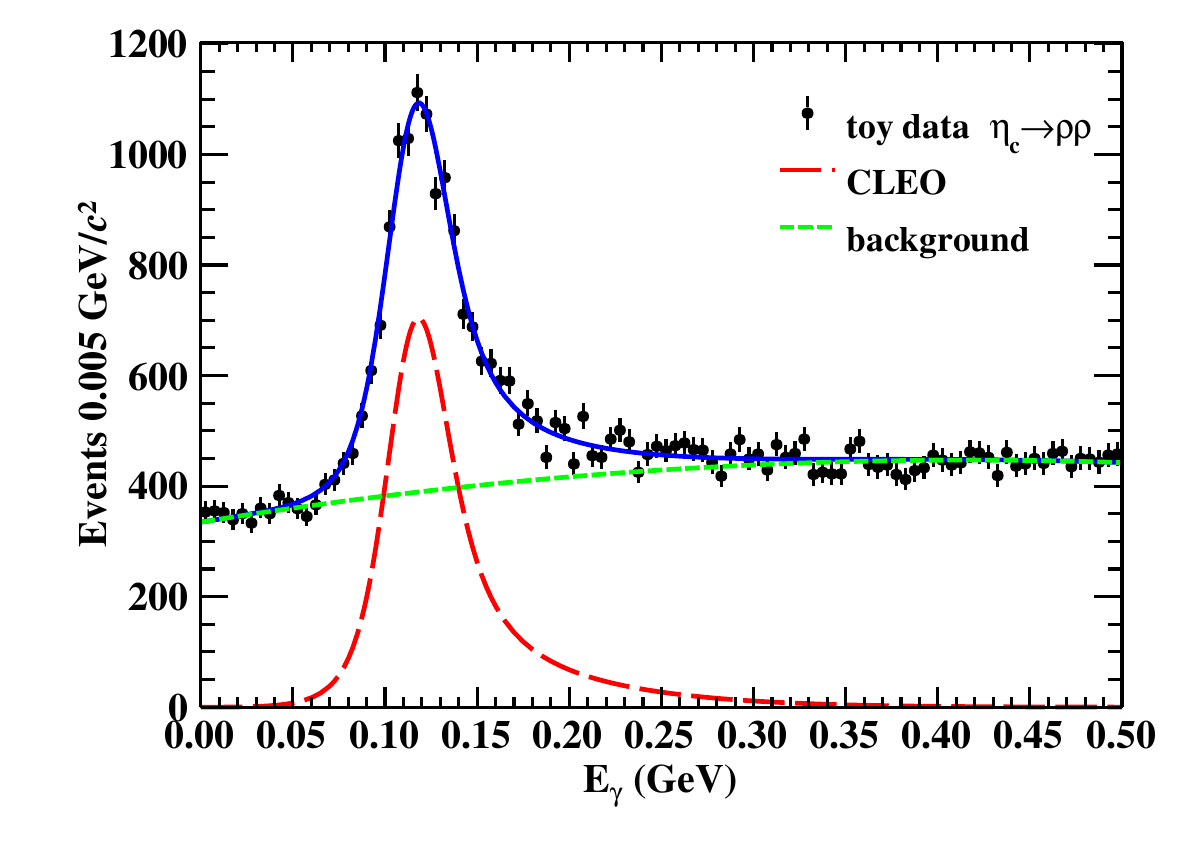}
\put(20,60){$(b)$}
\end{overpic}
\begin{overpic}[angle=0,width=0.32\textwidth]{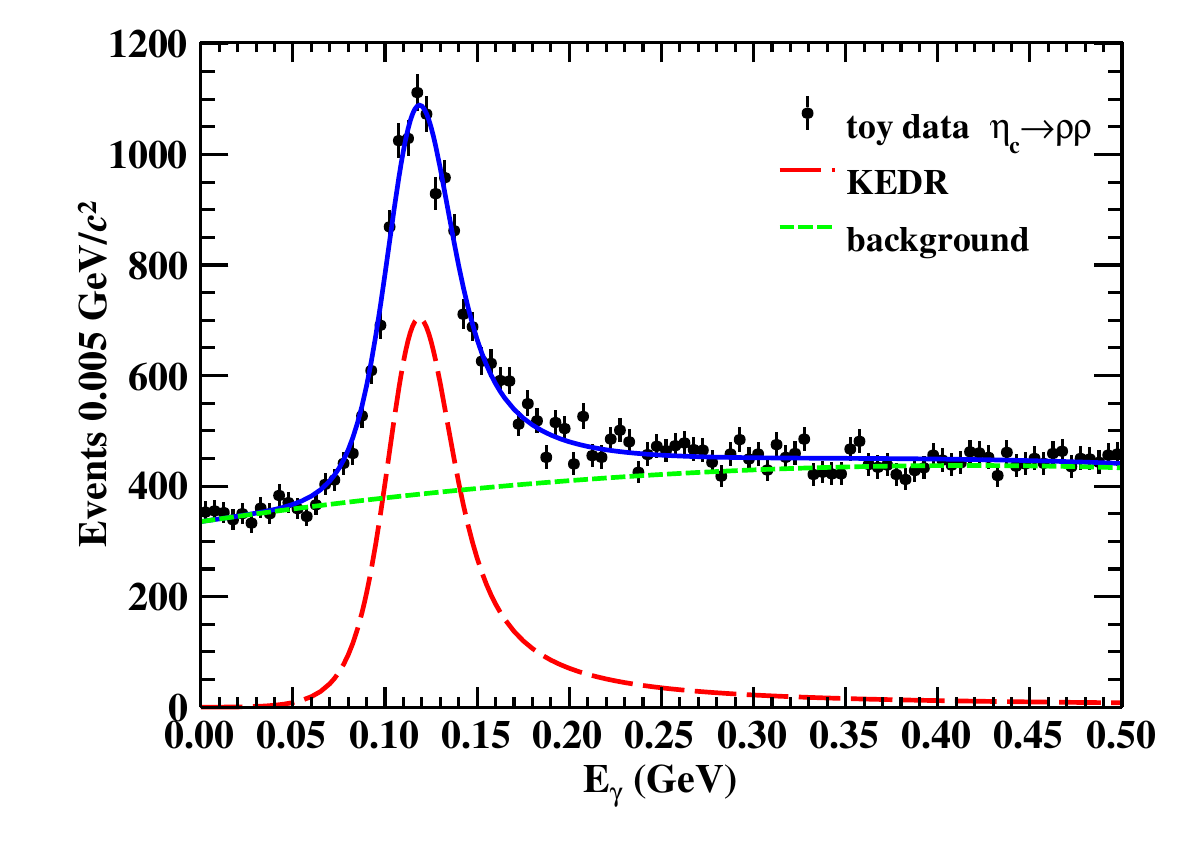}
\put(20,60){$(c)$}
\end{overpic}
}
\caption{The fit result to toy MC sample of $\eta_c \to \rr$ channel. Plot (a) shows the result with the damping
function obtained in this paper, plot (b) shows the result with CLEO's damping function, and (c) shows the result
of KEDR's damping function. The black dots are the toy MC sample, blue solid lines are the fit results, red dashed
lines are the signals with various damping functions, and green dashed lines are the backgrounds.}
\label{Fig:etac_var_fit}
\end{figure*}

We generate three toy MC samples to study the effect of new damping functions on the experimental measurements. Each
sample has $50,000$ events, including $15,000$ signal events of $\etac \to \rr$, $\XiXi$, or $\KKpiz$ decays, and
$35,000$ background events. The signal events are simulated using the line shape based on
Eq.~(\ref{eq:normal_line_shape}), in which the $BW$ function takes the world average values of the mass and the
width of $\etac$~\cite{PDG}, i.e., $m_{\etac} = 2984.1~\mevcs$ and $\Gamma_{\etac} = 30.5~\mev$. The background events
are simulated by a second order polynomial function with the coefficients chosen randomly. We fit these toy MC samples
by a combined function, in which the signal is described with a line shape according to
Eq.~(\ref{eq:normal_line_shape}), $D_{\rm CLEO}(\Eg)$, or $D_{\rm KDER}(\Eg)$, and the background is described by a
second-order polynomial function with floated parameters.  Figure~\ref{Fig:etac_var_fit} shows the example of $\etac \to \rr$ and the
fit results. Table~\ref{table:toydata_fit} summarizes all the fit results of the three
toy MC samples for $\etac \to \rr$, $\XiXi$, or $\KKpiz$ decays. It is obvious that different damping functions result in different resonant parameters of $\etac$,
and these differences vary according to $\etac$ decay modes. Compared to Eq.~(\ref{eq:normal_line_shape}), the
functions $D_{\rm CLEO}(\Eg)$ and $D_{\rm KDER}(\Eg)$ can yield smaller masses and larger widths for $\etac$.

\begin{table*}[!htbp]
\setlength{\abovecaptionskip}{0pt}
\setlength{\belowcaptionskip}{9pt}
\centering
\footnotesize  	% \tiny \scriptsize \footnotesize \small \normalsize \large \Large \LARGE \huge
\setlength{\tabcolsep}{20pt}					% column separation 12pt
\renewcommand{\arraystretch}{1.2} 	% row space
\caption{The obtained $\etac$ resonance parameters by fitting to the MC samples based on the line shape according to
our damping function. The mass and width of $\etac$ used to generate the toy MC samples are $2984.1~\mevcs$ and
$30.5~\mev$, respectively. The fitted results with damping functions from Eq.~(\ref{eq:normal_line_shape}),  CLEO,
and KEDR are compared.}
\begin{tabular}{c c c c c}
\hline \hline
Parameters\slash function   &  Eq.~(\ref{eq:normal_line_shape})  &CLEO   &KEDR \\
\hline
\multicolumn{4}{c}{$\etac \to \rr$ mode }\\
\hline
$m_{\etac}$ $(\mevcs)$    &	$2984.2 \pm 0.6$   &	$2982.2 \pm 0.5$  &	$2981.6 \pm 0.5$ \\
$\Gamma_{\etac}$ $(\mev)$  &	$29.8 \pm 1.3$  &	$30.8 \pm 1.4$    &	$32.6 \pm 1.5$ \\
\hline
\multicolumn{4}{c}{$\etac \to \XiXi$ mode }\\
\hline
$m_{\etac}$ $(\mevcs)$    & $2983.6 \pm 0.6$   &	$2981.6 \pm 0.5$   &	$2981.0 \pm 0.5$ \\
$\Gamma_{\etac}$ $(\mev)$  & $31.3 \pm 1.4$  & 	$31.5 \pm 1.4$    &		$33.1 \pm 1.5$ \\
\hline
\multicolumn{4}{c}{$\etac \to \KKpiz$ mode }\\
\hline
$m_{\etac}$ $(\mevcs)$    & $2984.5 \pm 0.5$   & $2982.8 \pm 0.5$   & $2982.2 \pm 0.5$ \\
$\Gamma_{\etac}$ $(\mev)$  & $30.0 \pm 1.2$  & $31.1 \pm 1.3$    & $33.1 \pm 1.4$ \\
\hline \hline
\end{tabular}
\label{table:toydata_fit}
\end{table*}

\begin{figure}[!htbp]
\centering
\includegraphics[angle=0,width=0.45\textwidth]{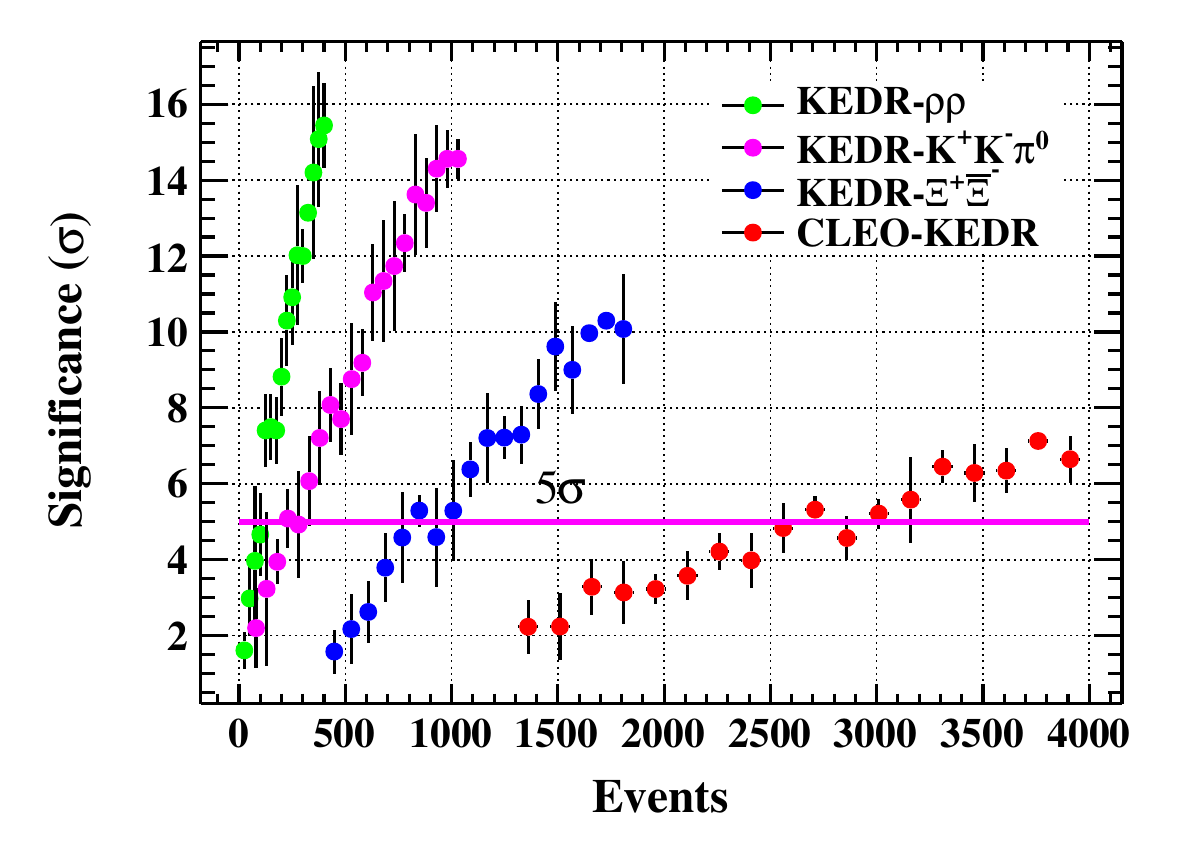}
\caption{The distinguishing significance between different damping functions concerning the number of signal
events. The horizontal pink solid line indicates the $5\sigma$ standard. The green dots indicate the significance
differing our damping function from the KEDR's with the $\eta_c \to \rr$ channel, the yank dots indicate the
significance differing our damping function from the KEDR's with the $\eta_c \to \KKpiz$ channel, the blue dots
indicate the significance of differing our damping function from the KEDR's with the $\eta_c \to \XiXi$ channel, and
the red dots indicate the significance of CLEO's damping function differing from that of KEDR's. }
\label{fig:sys_est}
\end{figure}

The ability to distinguish damping function hypotheses depends on the statistics of the data sample. To illustrate
this dependence, we calculate the significance of one damping function with respect to others, along with the number
of signal events. Since the background level is unknown, we do this calculation with ignoring the backgrounds. The
dependencies are displayed in Fig.~\ref{fig:sys_est}. The experimental models of CLEO and KEDR are more similar,
requiring at least $2,500$ signal events to reach $5\sigma$. To distinguish the new damping function from
$D_{\rm CLEO}(\Eg)$ and $D_{\rm KDER}(\Eg)$, we need only several hundreds of events. However, larger statistics
may be required if the background effect is considered.

\section{Summary and discussion}

We introduce two theoretically founded considerations to solve the problem of $\Egt$ divergence in the line shape of
the transition $\jpsi \to \gamma \etac$. They are the full-order contributions of the Bessel function in the overlap
integral of charmonium wave functions and the function of phase space, the second of which is usually ignored in
previous experimental measurements. It turns out that either can significantly suppress the divergent tail of the
$\jpsi \to \gamma \etac$ line shape, and a combination
of them effectively solves the divergent problem.

Taking into account the two considerations for future experimental measurements, we obtain the numerical damping function of the overlap integral of the $\jpsi$ and $\etac$ wave functions based on the non-relativistic potential model, as presented in Eq.~\ref{EqD}. Study with toy MC simulations show that combining this damping function with the phase space of specific $\etac$ decay channels, one could properly describe the line shape of $\jpsi \to \gamma \etac$, and precisely extract the mass and width of $\etac$. The toy MC study also shows that a few hundred signal events would be enough
to distinguish this new damping function from those adopted in previous measurements if the backgrounds were ignored. We recommend using the line shape obtained in this paper for the future $\jpsi \to \gamma \etac$ studies.

\clearpage
\newpage

\begin{acknowledgments}

This work is partly supported by the National Natural Key R\&D Program of China under Contract No. 2022YFA1601903, the National Natural Foundation of China (NSFC) under Contracts No. 12375083, No. 12275058 and No. 12175041.

\end{acknowledgments}

\bibliography{draft}

\end{document}